\documentstyle[aps,preprint,eqsecnum]{revtex}
\begin{document}
\author{Mirim Lee}
\address{TCSUH, Department of Physics, University of Houston,\\
Houston, TX 77204}
\author{James W. Dufty}
\address{Department of Physics, University of Florida, Gainesville, FL, 32611}
\title{Transport Far From Equilibrium --- Uniform Shear Flow}
\date{\today }
\draft
\maketitle

\begin{abstract}
The BGK model kinetic equation is applied to spatially inhomogeneous states
near steady uniform shear flow. The shear rate of the reference steady state
can be large so the states considered include those very far from
equilibrium. The single particle distribution function is calculated exactly
to first order in the deviations of the hydrodynamic field gradients from
their values in the reference state. The corresponding non-linear
hydrodynamic equations are obtained and the set of transport coefficients
are identified as explicit functions of the shear rate. The spectrum of the
linear hydrodynamic equations is studied in detail and qualitative
differences from the spectrum for equilibrium fluctuations are discussed.
Conditions for instabilities at long wavelengths are identified and
discussed.
\end{abstract}


\section{INTRODUCTION}

Nonequilibrium phenomena are well understood for states near equilibrium.
However, even the qualitative features of transport and fluctuations far
from equilibrium are poorly understood, due to the complexity of the
physical states and the lack of an adequate controlled theoretical
description in general. It is useful in this context to consider more
restrictive conditions and specialized states for which greater progress in
this difficult area can be made. Here we consider a low density simple
atomic gas for which transport properties are described by the non-linear
Boltzmann equation. For states near equilibrium, the Chapman-Enskog
expansion of the distribution function about the local equilibrium
distribution in terms of gradients of the hydrodynamic fields provides
approximations to a normal solution (one for which all space and time
dependence occurs through the hydrodynamic variables) \cite{ferziger}. From
this solution the corresponding hydrodynamic equations are obtained,
including explicit expressions for the associated transport coefficients. In
principle, this method applies to states far from equilibrium as well,
although calculation of the Chapman-Enskog expansion to higher orders in the
gradients is prohibitively difficult and questions of convergence remain
unresolved. An alternative approach is to expand in small gradients about a
more relevant reference state than local equilibrium. For example, consider
states near a homogeneous reference steady state. Deviations of the
hydrodynamic variables from their values in this state are characterized by
small relative spatial gradients. A modified Chapman-Enskog expansion can be
implemented to obtain the distribution function and hydrodynamic equations
to leading order in these gradients. Since the reference state can be far
from equilibrium the form of the hydrodynamic equations and the dependence
of the transport coefficients on parameters of the steady state will be
quite different from those for states near equilibrium.

In practice this program has not been carried out since determination of a
non-trivial reference steady state far from the equilibrium from the
Boltzmann equation is exceptionally rare \cite{santosrev}. In contrast,
exact results have been obtained for a number of physically interesting
special steady states \cite{duftyrev} using a model kinetic equation
designed to preserve the essential features of the Boltzmann equation while
admitting more practical analysis \cite{BGK}. In some cases these results
have been compared with those from Monte Carlo simulation far from
equilibrium, indicating that the kinetic model provides both a qualitative
and semi-quantitative representation of the underlying Boltzmann equation 
\cite{santoscomp}. The objective here is to use this kinetic model approach
for the special class of states at or near uniform shear flow. Uniform shear
flow is a particularly well-studied nonequilibrium state using numerical
methods (both Monte Carlo at low density \cite{santoscomp} and molecular
dynamics simulation at high density \cite{hoover,evans}), illustrating
rheological properties normally associated with complex molecular fluids\cite
{mdrev}. More importantly, it is one of the special cases for which an exact
solution to the model kinetic equation has been obtained \cite
{zwanzig,brey:91}, providing the necessary reference state for the modified
Chapman-Enskog expansion developed here. We look for solutions to the
kinetic equation as an expansion in small spatial gradients relative to an
exact solution for local uniform shear rather than around local equilibrium.
The heat and momentum fluxes are calculated from this solution to linear
order in these gradients, and a closed set of generalized hydrodynamic
equations is obtained. This is analogous to the non-linear Navier-Stokes
equations near equilibrium, except that here the reference state is
non-Maxwellian and a complex function of the shear rate. The associated
nonlinear transport coefficients are identified and examples calculated for
arbitrary values of the shear rate.

To expose the physical content of these hydrodynamic equations and their
differences from those for states near equilibrium, the hydrodynamic modes
are calculated from the associated linearized equations. In addition to
their dependence on the shear rate, these modes have a more complex
wavevector dependence than those from the Navier-Stokes equations due to the
broken symmetry of the reference state. More surprisingly, a new long
wavelength instability is found such that the hydrodynamic modes are growing
initially for any finite shear rate and sufficiently long wavelength \cite
{mirim:2}. A more detailed study of this instability and its verification
via computer simulation is described elsewhere \cite{instab}.

\section{KINETIC THEORY AND UNIFORM SHEAR FLOW}

\label{sec:kinetic}In this section the kinetic theory is defined, the
associated macroscopic conservation laws obtained, and the special solution
for steady uniform shear flow is described. In the next section, this
solution is generalized to a local reference state for an expansion to
describe a class of states near uniform shear flow and to obtain the
associated hydrodynamic equations.

Exact or even approximate solutions to the Boltzmann equation far from
equilibrium are exceptionally rare, due to the complexity of the nonlinear
collision operator. Therefore kinetic models have been introduced to replace
the Boltzmann collision operator with a simpler, more tractable operator.
The best studied of these is a single relaxation time model due to
Bhatnager, Gross and Krook (BGK model)\cite{BGK}. The essential qualities of
this model are its preservation of the exact equilibrium solution and all
five conservation laws. The BGK-Boltzmann kinetic equation is given by 
\begin{equation}
\left( \frac \partial {\partial t}+{\bf v}\cdot {\bf \nabla }_r\right)
f+m^{-1}{\bf \nabla }_v\cdot \left( {\bf F}_{ext}f\right) =-\nu (f-f_\ell ),
\label{BGKeq}
\end{equation}
where ${\bf F}_{ext}$ is an external force. The parameter $\nu $ in the (\ref
{BGKeq}) is a collision frequency which depends on an interaction law. This
frequency is a function of the density and temperature. At low density, it
can be written as 
\begin{equation}
\nu \sim n({\bf r},t)T^b({\bf r},t),  \label{freq}
\end{equation}
when the potential has a form $V(r)\sim r^{-l}$, with $b=1/2-2/l$. In the
case of Maxwell molecules ($l=4$), $b$ is zero so $\nu $ becomes independent
of temperature. For the hard sphere case $l\rightarrow \infty $ resulting in
a value for $b$ of $1/2$. Finally, $f_\ell ({\bf r},{\bf v},t)$ is the local
equilibrium distribution 
\begin{equation}
f_\ell ({\bf r},{\bf v},t)=n({\bf r},t)(\frac{\beta ({\bf r},t)m}{2\pi }%
)^{3/2}\exp \left( -\frac 12\beta ({\bf r},t)m({\bf v}-{\bf u}({\bf r}%
,t))^2\right),  \label{localf}
\end{equation}
where $n({\bf r},t)$, $T({\bf r},t)\equiv \left[ k_B\beta ({\bf r},t)\right]
^{-1}$, and ${\bf u}({\bf r},t)$ are the density, temperature, and flow
velocity of the nonequilibrium state. These hydrodynamic fields are defined
such that 
\begin{equation}
\int d{\bf v}\left( 
\begin{array}{c}
1 \\ 
{\bf v} \\ 
v^2
\end{array}
\right) \left( f({\bf r},{\bf v},t)-f_\ell ({\bf r},{\bf v},t)\right) \,=0,
\label{conserv}
\end{equation}
which assures that the BGK equation yields the correct conservation laws and
equilibrium stationary state in the absence of driving forces. More
explicitly (\ref{conserv}) gives 
\[
n({\bf r},t)\equiv \int d{\bf v}f({\bf r},{\bf v},t),\hspace{0.4in}n({\bf r}%
,t){\bf u}({\bf r},t)\equiv \int d{\bf v}\,{\bf v}f({\bf r},{\bf v},t),
\]
\begin{equation}
\frac 32n({\bf r},t)k_BT({\bf r},t)\equiv \int d{\bf v}\frac 12 m\,c^2
f({\bf r},{\bf v},t),  \label{conserv2}
\end{equation}
where ${\bf c} ={\bf v}-{\bf u}$.

The macroscopic conservation laws are obtained by taking moments of (\ref
{BGKeq}) and using the definitions (\ref{conserv2}) 
\begin{equation}
D_tn({\bf r},t)+n({\bf r},t)\nabla \cdot {\bf u}({\bf r},t)=0,  \label{eqn}
\end{equation}

\begin{equation}
D_tT({\bf r},t)+\frac 2{3k_Bn({\bf r},t)}[\nabla \cdot {\bf q}({\bf r}%
,t)+t_{ij}({\bf r},t)\partial _iu_j({\bf r},t)]=s({\bf r},t)  \label{eqT}
\end{equation}

\begin{equation}
D_tu_i({\bf r},t)+\left[ mn({\bf r},t)\right] ^{-1}\partial _jt_{ij}=0,
\label{equ}
\end{equation}
where $D_t\equiv \partial _t+{\bf u}\cdot \nabla $ is the material
derivative. The heat and momentum fluxes, ${\bf q}({\bf r},t)$ and $t_{ij}(%
{\bf r},t)$, are linear functionals of $f({\bf r},{\bf v},t)$ given by 
\begin{equation}
{\bf q}({\bf r},t)=\int d{\bf v}\frac 12m\,c^2{\bf c} f({\bf %
r},{\bf v},t),\hspace{0.4in}
t_{ij}({\bf r},t)=\int d{\bf v}\,m\,c_i
c_j f({\bf r},{\bf v},t).  \label{defq}
\end{equation}
The inhomogeneous term on the right side of the temperature equation, $s(%
{\bf r},t)$, is due to the external force ${\bf F}_{ext}$ , introduced to
serve as a thermostat. Several thermostats that have been used in both
theory and computer simulations. We choose here a force that is proportional
to the relative velocity ${\bf c}={\bf v}-{\bf u}({\bf r},t)$ 
\begin{equation}
{\bf F}_{ext}({\bf r},{\bf c},t)=-m\lambda (n({\bf r},t),T({\bf r}%
,t)){\bf c}.  \label{localF}
\end{equation}
The resulting source term $s({\bf r},t)$ in the equation for the temperature
becomes 
\begin{equation}
s({\bf r},t)=-2T({\bf r},t)\lambda (n({\bf r},t),T({\bf r},t)).
\label{source}
\end{equation}
The proportional ``constant'' $\lambda (n({\bf r},t),T({\bf r},t))$ is
determined by requiring stationarity of the system in the uniform shear flow
state (see below), and may depend on the local density, temperature, and
shear rate. In Appendix \ref{app:const} a different thermostat is considered
for comparison. The primary changes are the degree to which the external
force compensates for viscous heating away from the state of uniform shear
flow.

The fact that the fields are functionals of $f({\bf r},{\bf v},t)$ makes the
BGK-Boltzmann equation highly non-linear and difficult to solve in general.
However, in many cases an implicit solution can be given as an explicit
function of the velocity and functional of the fields. Then, the fields must
be determined self-consistently from the above macroscopic conservation
laws. One of the cases for which an exact solution is known is uniform shear
flow \cite{duftyrev,zwanzig,brey:91}. The uniform shear state is a planar
flow whose x-component of the flow velocity has a gradient along the y-axis, 
$u_{si}=a_{ij}r_j,\,\,\,\,a_{ij}=a\,\delta _{ix}\delta _{jy}$, where $a$ is
a constant shear rate. In addition, the density $n_s$, temperature $T_s$ ,
heat flux, and momentum flux are spatially constant. This state is generated
by a periodic boundary condition in the local Lagrangian frame. The viscous
heating induced by these boundary conditions is compensated by the external
force. It is easily verified that this macroscopic state is an exact
stationary solution to the above conservation laws (\ref{eqn}) - (\ref{equ})
if $\lambda (n({\bf r},t),T({\bf r},t))$ is chosen to be 
\begin{equation}
\lambda _s\equiv \lambda (n_s,T_s)=-\frac{a\,t_{xy,s}(a)}{3n_sk_BT_s}.
\label{deflam}
\end{equation}
Due to the simplicity of uniform shear flow state at the macroscopic level,
it has been studied extensively as a prototype of nonequilibrium states far
from equilibrium in theory \cite
{brey:91,dufty:83,dufty:86,rodrigues:2,dufty:79} and in computer simulations 
\cite{hoover,evans,erpenbeck,gomez}. To obtain the distribution function for
this state it is useful to express the kinetic equation (\ref{BGKeq}) in
terms of the velocity in the local rest frame, defined by $v_i^{\prime
}=v_i-a_{ij}r_j$. In this frame, the flow field vanishes and the macroscopic
state becomes spatially homogeneous. Consequently, we look for a stationary
solution to Eq.(\ref{BGKeq}) of the form $f({\bf r},{\bf v},t)=f_s({\bf v}%
^{\prime })$ 
\begin{equation}
L(v^{\prime },a)f_s({\bf v^{\prime }})=-\nu _sf_{s\ell }({\bf v^{\prime }}),%
\hspace{0.4in}L(v^{\prime },a)\equiv a_{ij}v_j^{\prime }\frac \partial {%
\partial v_i^{\prime }}+\lambda _s{\bf v}^{\prime }{\bf \cdot \nabla }%
_{v^{\prime }}+3\lambda _s-\nu _s.  \label{eqfs}
\end{equation}
The subscript $s$ denotes the stationary state value and $f_{s\ell }$ is the
corresponding local equilibrium distribution function with the hydrodynamic
fields for uniform shear flow. The solution to Eq.(\ref{eqfs}) is 
\begin{equation}
f_s({\bf v}^{\prime })=\nu _s\int_0^\infty d\tau \,e^{\tau L}f_\ell ({\bf v}%
^{\prime })=\nu _s\int_0^\infty dt\,e^{-t(\nu _s-3\lambda _s)}f_{s\ell
}(e^{\lambda _st}\Lambda _{ij}(-t)v_j^{\prime }).  \label{fs:1}
\end{equation}
The second equality follows from the property for an arbitrary function, $X(%
{\bf v})$ 
\begin{equation}
e^{tL}X({\bf v}^{\prime })=e^{(3\lambda _s-\nu _s)t}X(e^{\lambda _st}\Lambda
_{ij}(-t)v_j^{\prime }),\hspace{0.4in}\Lambda _{ij}(t)=\delta _{ij}-a_{ij}t.
\label{propL}
\end{equation}
where use has been made of $\exp (atv_x\partial _{v_x})X(v_x)=X(e^{at}v_x)$
and $\exp (atv_y\partial _{v_x})X(v_x)=X(v_x+atv_y)$. To determine $\lambda
_s$ as an explicit function of $a$, the component of the momentum flux $%
t_{xy,s}(a)$ can be calculated from Eqs.(\ref{defq}) and (\ref{fs:1}) to get
the self-consistent equation 
\begin{equation}
3\lambda _s(2\lambda _s+\nu _s)^2=\nu _sa^2.  \label{lamrel}
\end{equation}
This has one real solution and two complex conjugate solutions. The
physically relevant real value is 
\begin{equation}
\lambda _s(a)=\frac{2\,\nu _s}3\sinh [\cosh ^{-1}\frac 16(1+9\frac{a^2}{\nu
_s^2})]^2.  \label{lambda}
\end{equation}
With $\lambda _s$ known, the velocity distribution given by (\ref{fs:1}) is
completely determined.

Any transport property of interest now can be calculated by integration. A
detailed discussion can be found in \cite{brey:91} and comparison with Monte
Carlo simulations of the Boltzmann equation for shear flow is given in \cite
{gomez}. Only the transport properties associated with the heat and momentum
fluxes are considered further here. These can be calculated directly from (%
\ref{defq}) and (\ref{fs:1}) with the results 
\begin{equation}
{\bf q}(a)=0,\hspace{0.4in}t_{ij,s}(a)=(p_s+\frac 13a^2\Psi _1(a))\delta
_{ij}-\eta (a)(a_{ij}+a_{ji})-\Psi _1(a)a_{ik}a_{jk}  \label{tshear}
\end{equation}
Thus, the heat flux vanishes but the momentum flux describes non-trivial
rheological effects in terms of the hydrostatic pressure, $p=nk_BT$, the
shear viscosity, $\eta (a)\equiv -a^{-1}t_{xy}(a),$ and the viscometric
function, $\Psi _1(a)\equiv a^{-2}\left[ t_{yy}(a)-t_{xx}(a)\right] ,$ where 
\begin{eqnarray}
\eta (a)=\frac{\nu _s}{(2\lambda _s(a)+\nu _s)^2}p_s,\hspace{0.4in}\Psi
_1(a)=-\frac{6\lambda _s(a)}{a^2(2\lambda _s(a)+\nu _s)}p_s.  \label{etac}
\end{eqnarray}
In general there is a second independent viscometric function, $\,\Psi
_2(a)\equiv a^{-2}\left[ t_{zz}(a)-t_{yy}(a)\right] $, which vanishes for
our kinetic model. The magnitudes of these transport coefficients are
monotonically decreasing functions of the shear rate, and have been
discussed in detail elsewhere \cite{brey:91}.

\section{HYDRODYNAMICS NEAR UNIFORM SHEAR FLOW}

\label{sec:hydro} In this section we consider states that deviate from
uniform shear flow by small spatial gradients. A solution to the
BGK-Boltzmann equation (\ref{BGKeq}) is obtained by a variant of the
Chapman-Enskog method whereby the distribution function is expanded about a 
{\em local} uniform shear flow reference state in terms of the small spatial
gradients of the hydrodynamic fields relative to those of uniform shear
flow. This is analogous to the usual Chapman-Enskog expansion about a local
equilibrium distribution. The solution obtained in this way can be used to
calculate the heat and momentum fluxes in terms of the hydrodynamic
variables, so that Eqs. (\ref{eqn})-(\ref{equ}) become a closed set of
hydrodynamic equations. The analysis here is carried out to first order in
the gradients. For small shear rate the usual Navier Stokes results are
recovered, where the heat flux is given by Fourier's law and the momentum
flux is given by Newton's viscosity law. However, for large shear rates
these fluxes and the corresponding hydrodynamic equations are more complex.

To construct the Chapman-Enskog expansion we look for solutions of the form 
\begin{equation}
f({\bf r},{\bf v},t)=f({\bf v}^{\prime}, y_\alpha ({\bf r},t)),
\label{normal}
\end{equation}
where $y_\alpha ({\bf r},t)$ are the hydrodynamic fields, and $v_i^{\prime
}=v_i-a_{ij}r_j$. This representation expresses the fact that the space
dependence of the reference shear flow is completely absorbed in the
relative velocity variable, ${\bf v}^{\prime }$, and all other space and
time dependence occurs entirely through a functional dependence on the
hydrodynamic variables, $y_\alpha ({\bf r},t)$. This is an example of a
``normal'' solution, which is expected to result from a wide class of
initial conditions at long times and large space scales. It is essential for
a hydrodynamic description since the velocity average of any quantity
becomes a functional of the hydrodynamic fields. Approximate solutions to
the BGK equation are obtained by expanding (\ref{normal}) in a formal
uniformity parameter, $\epsilon $, that measures the spatial gradients of
the fields $y_\alpha ({\bf r},t)$ 
\begin{equation}
f({\bf v}^{\prime},y_\alpha ({\bf r},t))=f^{(0)}({\bf v}^{\prime },y_\alpha
({\bf r},t))+\epsilon f^{(1)}({\bf v}^{\prime },y_\alpha ({\bf r},t))+\cdots
.  \label{expf}
\end{equation}
This expansion leads to a corresponding expansion for the heat and momentum
fluxes when substituted into (\ref{defq}) 
\begin{equation}
t_{ij}=t_{ij}^{(0)}+\epsilon \,t_{ij}^{(1)}+\cdot \cdot \cdot ,\hspace{0.4in}%
{\bf q}={\bf q}^{(0)}+\epsilon \,{\bf q}^{(1)}+\cdot \cdot \cdot ,
\label{expflux}
\end{equation}
\begin{equation}
t_{ij}^{(r)}=\int d{\bf v}m\,c_i c_j f^{(r)},\hspace{0.3in}%
{\bf q}^{(r)}=\int d{\bf v}\frac 12m\,c^{2}{\bf c} f^{(r)}.
\label{defflux}
\end{equation}
Finally, use of this expansion for the fluxes in the conservation laws (\ref
{eqn})-(\ref{equ}) identifies an expansion for the time derivatives of the
fields in powers of the uniformity parameter 
\begin{equation}
\frac \partial {\partial t}=\frac{\partial ^{(0)}}{\partial t}+\epsilon 
\frac{\partial ^{(1)}}{\partial t}+\cdots .  \label{time}
\end{equation}
These results provide the basis for generating the Chapman-Enskog solution
to the BGK-Boltzmann equation.

The BGK-Boltzmann equation in terms of the variable ${\bf v}^{\prime }$ is
obtained directly from (\ref{BGKeq}) 
\begin{equation}
\left( \frac \partial {\partial t}+(v_i^{\prime }+a_{ij}r_j)\frac \partial {%
\partial r_i}-L(v^{\prime },a)+\lambda \delta {\bf u}\cdot {\bf \nabla }_{%
{\bf v}^{\prime }}\right) f({\bf r},{\bf v}^{\prime },t)=\nu f_\ell ({\bf r},%
{\bf v}^{\prime },t),  \label{BGKref}
\end{equation}
where $\delta {\bf u}={\bf u}-{\bf u}_s$. Also, the operator $L(v^{\prime
},a)$ is defined by (\ref{eqfs}) except with $\nu _s,\lambda _s$ replaced by 
$\nu ,\lambda $ as functions of $n({\bf r},t)$ and $T({\bf r},t)$. The form
of $\lambda $ is still to be determined at this point. The spatial gradient
in the second term on the left side is taken at constant ${\bf v}^{\prime }$%
, and consequently this term is of first order in the uniformity parameter.
Substituting the expansions (\ref{expf}) and (\ref{time}) into (\ref{BGKref}%
) and equating coefficients of each degree in the uniformity parameter leads
to the equations for $f^{(0)}$ and $f^{(1)}$ 
\begin{equation}
\left( \frac{\partial ^{(0)}}{\partial t}-L(v^{\prime },a)+\lambda \delta 
{\bf u}\cdot {\bf \nabla }_{{\bf v}^{\prime }}\right) f^{(0)}=\nu f_\ell ,
\label{zeroth}
\end{equation}
\begin{equation}
\left( \frac{\partial ^{(0)}}{\partial t}-L(v^{\prime },a)+\lambda \delta 
{\bf u\cdot \nabla }_{{\bf v}^{\prime }}\right) f^{(1)}=-\left( \frac{%
\partial ^{(1)}}{\partial t}+(v_i^{\prime }+a_{ij}r_j)\frac \partial {%
\partial r_i}\right) f^{(0)}.  \label{first}
\end{equation}
To lowest order in the expansion the conservation laws give 
\begin{equation}
\frac{\partial ^{(0)}n}{\partial t}=0,\hspace{0.3in}\frac{\partial
^{(0)}\delta u_i}{\partial t}+a_{ij}\delta u_j=0,\hspace{0.3in}\frac 32nk_B%
\frac{\partial ^{(0)}T}{\partial t}+a_{ij}t_{ij}^{(0)}+3\lambda nk_BT=0,
\label{condi}
\end{equation}
where $t_{ij}^{(0)}$ is defined by (\ref{defflux}). The parameter $\lambda $
of the external force is now chosen to impose $\frac{\partial ^{(0)}T}{%
\partial t}=0$, i.e., 
\begin{equation}
\lambda (n({\bf r},t),T({\bf r},t))=-at_{xy}^{(0)}({\bf r},t)/3n({\bf r}%
,t)k_BT({\bf r},t).  \label{locallam}
\end{equation}
The solution to (\ref{zeroth}) is obtained in a way similar to that for (\ref
{eqfs}) with the result 
\begin{eqnarray}
&&f^{(0)}({\bf v}^{\prime },y_\alpha ({\bf r},t))=\nu ({\bf r},t)n({\bf r}%
,t)\left( \frac m{2\pi k_BT({\bf r},t)}\right) ^{3/2}\int_0^\infty d\tau
\,e^{-\nu ({\bf r},t)\tau }e^{3\lambda ({\bf r},t)\tau }  \nonumber \\
&&\quad \quad \quad \exp \left( -\frac m{2k_BT({\bf r},t)}e^{2\lambda ({\bf r%
},t)\tau }[\Lambda _{ij}(-\tau )(v_j^{\prime }-\delta u_j({\bf r}%
,t))]^2\right) ,  \label{f0}
\end{eqnarray}
where $\Lambda _{ij}(t)$ is defined in (\ref{propL}). The relationship of (%
\ref{f0}) to (\ref{fs:1}) is analogous to the relationship of the local
equilibrium distribution to the strict equilibrium distribution, where the
former is obtained from the latter by replacing the hydrodynamic fields with
their actual nonequilibrium values. To determine $\lambda ({\bf r},t)\equiv
\lambda (n({\bf r},t),T({\bf r},t))$, the momentum flux $t_{xy}^{(0)}({\bf r}%
,t)$ is calculated using (\ref{defflux}) and (\ref{f0}) to obtain 
\begin{equation}
t_{xy}^{(0)}({\bf r},t)=\frac{-a\,\nu ({\bf r},t)}{\left( 2\lambda ({\bf r}%
,t)+\nu ({\bf r},t)\right) ^2}\,n({\bf r},t)k_BT({\bf r},t),  \label{txy:1}
\end{equation}
where $\nu ({\bf r},t)\equiv \nu (n({\bf r},t),T({\bf r},t))$. Use of (\ref
{locallam}) then gives finally 
\begin{equation}
3\lambda ({\bf r},t)(2\lambda ({\bf r},t)+\nu ({\bf r},t))^2=\nu ({\bf r}%
,t)a^2.  \label{lamrel:2}
\end{equation}
This shows that $\lambda (n({\bf r},t),T({\bf r},t))$ is the same as (\ref
{lamrel}) for uniform shear flow, except that the density and temperature
are replaced by their values for the general nonequilibrium state. With this
result for $\lambda ({\bf r},t)$ the solution (\ref{f0}) is completely
determined.

Next, consider the solution to (\ref{first}) for the contributions to first
order in the spatial gradients. As shown in Appendix \ref{app:X}, the right
side is a linear combination of the hydrodynamic gradients. Consequently, $%
f^{(1)}$ necessarily has the same form 
\begin{equation}
f^{(1)}({\bf v}^{\prime },y_\alpha ({\bf r},t))=X_{n,i}\frac{\partial n}{%
\partial r_i}+X_{T,i}\frac{\partial T}{\partial r_i}+X_{u_k,i}\frac{\partial
\delta u_k}{\partial r_i},  \label{f1form}
\end{equation}
where the coefficients $X_{\beta ,i}(y_\alpha ({\bf r},t),{\bf v}^{\prime })$
are functions of the velocity to be determined from substitution of (\ref
{f1form}) into (\ref{first}) 
\begin{equation}
\left( \frac{\partial ^{(0)}}{\partial t}-L(v^{\prime },a)+\lambda \,\delta 
{\bf u\cdot \nabla }_{{\bf v}^{\prime }}\right) X_{\beta
,k}-aX_{u_x,k}\delta _{\beta ,u_y}=-Y_{\beta ,k}.  \label{Xeq}
\end{equation}
Here $\beta $ denotes $n,T,u_x,u_y,$ and $u_z$, and $i,k$ represents
Cartesian coordinates. The last term on the left side originates from $\frac{%
\partial ^{(0)}}{\partial t}\frac{\partial \delta u_k}{\partial r_i}=\frac 
\partial {\partial r_i}\frac{\partial ^{(0)}}{\partial t}\delta u_k$ and (%
\ref{condi}). The functions $Y_{\beta ,k}$ are given explicitly in terms of $%
f^{(0)}$ and the flux $t_{ij}^{(1)}$, which must be determined
self-consistently from $f^{(1)}$. The procedure is to solve (\ref{Xeq}) to
find $f^{(1)}$ in terms of $t_{ij}^{(1)}$ and then to use this result to
derived a self-consistent equation for $t_{ij}^{(1)}$. Further details can
be found in Appendix \ref{app:X} along with the explicit solution to (\ref
{Xeq}).

In this way the complete normal solution near the uniform shear state is
constructed to order $\epsilon $. The heat and momentum fluxes then are
calculated from (\ref{defflux}). The first terms, $t_{ij}^{(0)}(a)$ and $%
q_i^{(0)}(a)$, represent the transport properties of the{\em \ local}
reference state. They are the same as those of the previous section, (\ref
{tshear}), except that the density and temperature are replaced by their
values for the general nonequilibrium state. An important consequence of
this replacement is that their gradients are no longer zero and thus they
now contribute to the hydrodynamic equations. The second terms, $%
t_{ij}^{(1)}(a)$ and $q_i^{(1)}(a)$ provide new transport coefficients
describing dissipation due to spatial deviations uniform shear flow 
\begin{equation}
q_i^{(1)}(a)=-\left( \xi _{T,j}^i(a)\frac{\partial T}{\partial r_j}+\xi
_{n,j}^i(a)\frac{\partial n}{\partial r_j}\right) ,\hspace{0.3in}%
t_{ij}^{(1)}(a)=-\gamma _{lk}^{ij}(a)\frac{\partial \delta u_l}{\partial r_k}%
.  \label{newtij}
\end{equation}
It is understood that $\nu $, $\lambda (a)$, $p$, and the new transport
coefficients all are functions of the local nonequilibrium temperature and
density, as well as the shear rate $a$. The method for calculating $\gamma
_{lk}^{ij}(a)$and $\xi _{\alpha ,j}^i(a)$ and some detailed forms of the
coefficients are given in Appendix \ref{app:coefs}; a more complete listing
can be obtained on request from the authors. Since the reference state is
anisotropic there are new transport coefficients, reflecting the broken
fluid symmetry, which do not exist for hydrodynamics near equilibrium.
Fourier's law for the heat flux is modified by a thermal conductivity
tensor, $\xi _{T,j}^i(a)$, which has the form 
\begin{equation}
\xi _{T,j}^i(a)=\xi _T^{(1)}(a)\delta _{ij}+\xi _T^{(2)}(a)a_{ij}+\xi
_T^{(3)}(a)a_{ji}+\xi _T^{(4)}(a)a_{ik}a_{jk}+\xi _T^{(5)}(a)a_{ki}a_{kj}.
\label{thermal}
\end{equation}
characterized by five scalar ``thermal conductivities'' characterizing this
tensor. The anisotropy is in the $x,y$ plane so, for example, a temperature
gradient in the $x-$direction leads to a heat flux in both $x-$ and $y-$%
directions. At zero shear rate only the contribution from $\xi _T^{(1)}(a)$
survives with $\xi _T^{(1)}(0)=\kappa $, where $\kappa =5k_Bp/2m\nu $ is the
thermal conductivity for the BGK model. An additional difference from
Fourier's law is a contribution to the heat flux from a density gradient
characterized by the transport tensor, $\xi _{n,j}^i(a)$. This has a
representation like (\ref{thermal}) in terms of five additional scalar
transport coefficients. The asymptotic behavior for small shear rates is $%
\xi _{n,j}^i(a)\rightarrow \left( a_{ij}+a_{ji}\right) m\left( \beta /\nu
\right) ^2$ which vanishes for zero shear rate as expected. The momentum
flux is expressed in terms of a viscosity tensor, $\gamma _{lk}^{ij}(a)$, of
degree four which is symmetric and traceless in $ij$. There are $19$
independent viscosity coefficients. At zero shear rate $\gamma _{lk}^{ij}(a)$
reduces to Newton's viscosity law, $\gamma _{lm}^{ij}(a=0)=\eta (\delta
_{i\ell }\delta _{jm}+\delta _{im}\delta _{j\ell }-\frac 23\delta
_{ij}\delta _{\ell m})$, where $\eta =p/\nu $ is the shear viscosity for the
BGK-model.

The corresponding non-linear hydrodynamic equations are obtained directly
from these results and the conservation laws 
\begin{equation}
D_tn+n{\bf \nabla }\cdot \delta {\bf u}=0,  \label{eq:1}
\end{equation}
\begin{eqnarray}
\frac 32nk_BD_tT &-&a_{ij}\gamma _{lm}^{ij}\frac{\partial \delta u_l}{%
\partial r_m}+\frac{\partial \delta u_i}{\partial r_j}[t_{ij}^{(0)}-\gamma
_{lm}^{ij}\frac{\partial \delta u_l}{\partial r_m}]  \nonumber \\
&&-\frac \partial {\partial r_i}[\xi _{n,j}^i\frac{\partial n}{\partial r_j}%
+\xi _{T,j}^i\frac{\partial T}{\partial r_j}]=0,  \label{eq:2}
\end{eqnarray}
\begin{equation}
D_t\delta u_k+\rho ^{-1}\frac \partial {\partial r_i}[t_{ik}^{(0)}-\gamma
_{lm}^{ik}\frac{\partial \delta u_l}{\partial r_m}]+a_{kj}\delta u_j=0,
\label{eq:3}
\end{equation}
where $D_t=\partial _t+{\bf u\cdot \nabla }$. This is a primary result of
our analysis. These equations are analogous to the non-linear Navier-Stokes
equations for a fluid near local equilibrium, and reduce to them for zero
shear rate. More generally, the reference state is local shear flow which
can be very far from equilibrium. Furthermore, there is no restriction on
the deviations from this reference state, $\delta y_a$, since the non-linear
dependence of all coefficients on $y_a({\bf r},t)$ has been retained. The
spatial gradients relative to shear flow must be small, however, and the
equations are accurate to second order in these gradients. The terms
proportional to $\gamma _{lm}^{ij}$ and $t_{ij}^{(0)}$ in (\ref{eq:2})
represent viscous heating due to the excess gradient of the flow velocity
relative to the reference state. There is no viscous heating from the
reference state alone, since the local thermostat has been chosen to cancel
it, but for states near uniform shear flow the thermostat cannot compensate
for effects due to the gradients relative to the reference state. The
implications of this new hydrodynamic description far from equilibrium are
elaborated in the next section by an analysis of the associated linear
hydrodynamic modes.

\section{HYDRODYNAMIC MODES}

\label{sec:correl} The above hydrodynamic equations are restricted to small
spatial gradients relative to the reference state of uniform shear flow. If
in addition the initial perturbations $\delta y_a(0)$ are small then these
equations can be linearized with respect to $\delta y_a(t)$. The resulting
set of five linear equations defines the hydrodynamic modes, or linear
response excitations to small perturbations. If all of these modes decay in
time the state is linearly stable. Otherwise, a growth of these modes
signals an onset of instability that is ultimately controlled by the
dominance of nonlinear terms. In this section we determine the hydrodynamic
modes for states far from equilibrium and contrast them with those for
states near equilibrium. 

The linearized hydrodynamic equations follow directly from (\ref{eq:1}) - (%
\ref{eq:3}) 
\begin{equation}
\left( \frac \partial {\partial t}+a_{ij}r_j\frac \partial {\partial r_i}%
\right) \,\delta n+n_s{\bf \nabla }\cdot \delta {\bf u}=0,  \label{linear:1}
\end{equation}
\begin{equation}
\left( \frac \partial {\partial t}+a_{ij}r_j\frac \partial {\partial r_i}%
\right) \,\delta T+\frac 2{3n_sk_B}\left( -a\gamma
_{ij,s}^{xy}+t_{ij,s}^{(0)}\right) \frac{\partial \delta u_i}{\partial r_j}-%
\frac 2{3n_sk_B}\left[ \xi _{nj,s}^i\frac{\partial ^2\delta n}{\partial
r_i\partial r_j}+\xi _{Tj,s}^i\frac{\partial ^2\delta T}{\partial
r_i\partial r_j}\right] =0,  \label{linear:2}
\end{equation}
\begin{equation}
\left( \frac \partial {\partial t}+a_{ij}r_j\frac \partial {\partial r_i}%
\right) \delta u_k+\rho _s^{-1}\left[ \left( \frac{\partial t_{ik}^{(0)}}{%
\partial n}\right) _s\frac{\partial \delta n}{\partial r_i}+\left( \frac{%
\partial t_{ik}^{(0)}}{\partial T}\right) _s\frac{\partial \delta T}{%
\partial r_i}-\gamma _{lm,s}^{ik}\frac{\partial ^2\delta u_l}{\partial
r_i\partial r_m}\right] +a_{kj}\delta u_j=0,  \label{linear:3}
\end{equation}
To analyze these equations it is convenient to transform to the local
Lagrangian frame, $r_i^{\prime }=r_i-u_{si}(r)t=\Lambda _{ij}(t)r_j$. The
Lees-Edwards boundary conditions then become simple periodic boundary
conditions in the variable ${\bf r}^{\prime }$. A Fourier representation is
defined by, 
\begin{equation}
\delta \widetilde{y}_\alpha ({\bf k},t)=\int d{\bf r}^{\prime }e^{i{\bf %
k\cdot r}^{\prime }}\delta y_\alpha ({\bf r},t)=\int d{\bf r}e^{i{\bf k}(t)%
{\bf \cdot r}}\delta y_\alpha ({\bf r},t),  \label{Fourier}
\end{equation}
where the periodicity requires $k_i=2n_i\pi /L_i$, where $n_i$ are integers
and $L_i$ are the linear dimensions of the system. In the second equality, $%
{\bf k}(t)$, is given by 
\begin{equation}
k_i(t)=k_j\Lambda _{ji}(t).  \label{k(t)}
\end{equation}
The linearized hydrodynamic equations in this Fourier representation are, 
\begin{equation}
\frac \partial {\partial t}\delta \widetilde{y}_\alpha +(A_{\alpha \nu
}-ik_j(t)B_{\alpha \nu ,j}+k_j(t)k_l(t)D_{\alpha \nu ,jl})\delta \widetilde{y%
}_\nu =0,  \label{generaleq}
\end{equation}
where in addition the dependent variables now have been scaled to
dimensionless forms, 
\begin{equation}
\delta {\widetilde{y}}_\alpha =\left\{ \frac{\delta \widetilde{n}}{n_s},\,%
\sqrt{\frac 32}\frac{\delta \widetilde{T}}{T_s},\,\sqrt{\frac m{k_BT_s}}%
\delta \widetilde{{\bf u}}\right\}.   \label{yalpha}
\end{equation}
A summation convention applies and Latin indices denote Cartesian
coordinates. The three matrices $A_{\alpha \beta },B_{\alpha \beta
},D_{\alpha \beta }$ are 
\begin{equation}
A_{\alpha \beta }=a\,\delta _{\alpha 3}\delta _{\beta 4},  \label{matA}
\end{equation}
\begin{equation}
B_{\alpha \beta j}=\sqrt{\frac{k_BT_s}m}\left( 
\begin{array}{ccccc}
0 & 0 & \delta _{jx} & \delta _{jy} & \delta _{jz} \\ 
0 & 0 & B_{2xj} & B_{2yj} & B_{2zj} \\ 
\frac n{p_s}\left( \frac{\partial t_{jx}^{(0)}}{\partial n}\right) _s & 
\sqrt{\frac 23}\frac{T_s}{p_s}\left( \frac{\partial t_{jx}^{(0)}}{\partial T}%
\right) _s & 0 & 0 & 0 \\ 
\frac n{p_s}\left( \frac{\partial t_{jy}^{(0)}}{\partial n}\right) _s & 
\sqrt{\frac 23}\frac{T_s}{p_s}\left( \frac{\partial t_{jy}^{(0)}}{\partial T}%
\right) _s & 0 & 0 & 0 \\ 
\frac n{p_s}\left( \frac{\partial t_{jz}^{(0)}}{\partial n}\right) _s & 
\sqrt{\frac 23}\frac{T_s}{p_s}\left( \frac{\partial t_{jz}^{(0)}}{\partial T}%
\right) _s & 0 & 0 & 0
\end{array}
\right) ,  \label{matB}
\end{equation}
where $B_{2ij}=\sqrt{\frac 23}\frac 1{p_s}(-a\gamma
_{ij,s}^{xy}+t_{ij,s}^{(0)})$. The matrix $D_{\alpha \beta }$ is

\begin{equation}
D_{\alpha \beta jl}=\left( 
\begin{array}{ccccc}
0 & 0 & 0 & 0 & 0 \\ 
\sqrt{\frac 23}\frac{n_s}{p_s}\xi _{nl,s}^j & \frac 23\frac{T_s}{p_s}\xi
_{Tl,s}^j & 0 & 0 & 0 \\ 
0 & 0 & \rho ^{-1}\gamma _{xl,s}^{jx} & \rho ^{-1}\gamma _{yl,s}^{jx} & \rho
^{-1}\gamma _{zl,s}^{jx} \\ 
0 & 0 & \rho ^{-1}\gamma _{xl,s}^{jy} & \rho ^{-1}\gamma _{yl,s}^{jy} & \rho
^{-1}\gamma _{zl,s}^{jy} \\ 
0 & 0 & \rho ^{-1}\gamma _{xl,s}^{jz} & \rho ^{-1}\gamma _{yl,s}^{jz} & \rho
^{-1}\gamma _{zl,s}^{jz}
\end{array}
\right) .  \label{matD}
\end{equation}
The homogeneous solution to equations (\ref{generaleq}) can be calculated
easily by putting ${\bf k}=0$, 
\begin{equation}
\delta {\widetilde{y}}_\alpha (t)=\left[ e^{-At}\right] _{\alpha \beta
}\delta {\widetilde{y}}_\beta (0)=\left[ 1-At\right] _{\alpha \beta }\delta {%
\widetilde{y}}_\beta (0).
\end{equation}
The second equality follows from the property $A^2=0$. Consequently, all
fields are constant except $\delta u_x$ which behaves as, 
\begin{equation}
\delta \widetilde{u}_x(t)=\widetilde{u}_x(0)-at\delta \widetilde{u}_y(0).
\end{equation}
The homogeneous state is unstable to an initial perturbation in $\delta 
\widetilde{u}_y$, leading to unbounded linear change in time. Stability is
still possible at finite $k$ if this behavior is modulated by exponential
hydrodynamic damping factors $\sim e^{-\alpha k^2t}$ with $\alpha > 0$.

To simplify the analysis at ${\bf k\neq 0}$ we allow the perturbation only
along the velocity gradient direction, i.e. ${\bf k=}k\,\widehat{{\bf y}}$.
In this case the linear hydrodynamic equations have time-independent
coefficients (i.e., ${\bf k}(t)={\bf k)}$, 
\begin{equation}
\left( \frac \partial {\partial t}+F\right) _{\alpha \nu }\delta \widetilde{y%
}_\nu =0,\hspace{0.4in}F_{\alpha \beta }=A_{\alpha \beta }-ik\,B_{\alpha
\beta }+k^2\,D_{\alpha \beta },  \label{myeq}
\end{equation}
and the matrices $B$ and $D$ now take the simpler forms, 
\begin{equation}
B_{\alpha \beta }=\sqrt{\frac{k_BT_s}m}\left( 
\begin{array}{ccccc}
0 & 0 & 0 & 1 & 0 \\ 
0 & 0 & B_{2xy} & B_{2yy} & 0 \\ 
\frac{n_s}{p_s}\left( \frac{\partial t_{yx}^{(0)}}{\partial n}\right) _s & 
\sqrt{\frac 23}\frac{T_s}{p_s}\left( \frac{\partial t_{yx}^{(0)}}{\partial T}%
\right) _s & 0 & 0 & 0 \\ 
\frac{n_s}{p_s}\left( \frac{\partial t_{yy}^{(0)}}{\partial n}\right) _s & 
\sqrt{\frac 23}\frac{T_s}{p_s}\left( \frac{\partial t_{yy}^{(0)}}{\partial T}%
\right) _s & 0 & 0 & 0 \\ 
0 & 0 & 0 & 0 & 0
\end{array}
\right) ,  \label{matbky}
\end{equation}

\begin{equation}
D_{\alpha \beta }=\left( 
\begin{array}{ccccc}
0 & 0 & 0 & 0 & 0 \\ 
\sqrt{\frac 23}\frac{n_s}{p_s}\xi _{ny,s}^y & \frac 23\frac{T_s}{p_s}\xi
_{Ty,s}^y & 0 & 0 & 0 \\ 
0 & 0 & \rho _s^{-1}\gamma _{xy,s}^{yx} & \rho _s^{-1}\gamma _{yy,s}^{yx} & 0
\\ 
0 & 0 & \rho _s^{-1}\gamma _{xy,s}^{yy} & \rho _s^{-1}\gamma _{yy,s}^{yy} & 0
\\ 
0 & 0 & 0 & 0 & \rho _s^{-1}\gamma _{zy,s}^{yz}
\end{array}
\right) .  \label{matdky}
\end{equation}
Equation (\ref{myeq}) can be solved by Laplace transformation, 
\begin{equation}
\widehat{\delta y}_\alpha ({\bf k},z)=\int_0^\infty dt\,e^{-tz}\delta 
\widetilde{y}_\alpha ({\bf k},t)=[zI+F({\bf k},z)]_{\alpha \nu }^{-1}\delta 
\widetilde{y}_\nu ({\bf k},t=0).  \label{lapdef}
\end{equation}
The eigenvalues $\omega ^{(i)}({\bf k},a)$ of the matrix $F({\bf k},z)$
define the five simple hydrodynamic poles at $z=-\omega ^{(i)}({\bf k},a)$,
which determine the dominant dynamics of the $\delta \widetilde{y}_\alpha (%
{\bf k},t)$ at large $t$ and small $k$. At equilibrium ($a=0$), the
hydrodynamic modes of the Navier-Stokes equations are recovered (two sound
modes, a heat mode and a two-fold degenerate shear mode) for long
wavelengths ($k\rightarrow 0$), 
\begin{equation}
\omega ^{(i)}({\bf k},0)\rightarrow \omega _{NS}^{(i)}(k)=\left( 
\begin{array}{c}
ick+\Gamma k^2 \\ 
-ick+\Gamma k^2 \\ 
D_Tk^2 \\ 
\left( \eta /\rho \right) k^2 \\ 
\left( \eta /\rho \right) k^2
\end{array}
\right) ,  \label{mode1}
\end{equation}
where $c=\sqrt{5/3\beta m}$ is the sound velocity, 
$\Gamma =D_{T}/3+\frac{2\eta}{3\rho}=1/\beta m\nu $
is the sound damping constant, 
$D_T=\xi^{y}_{Ty,s}(a=0)/\rho C_p=1/\beta m\nu $ is the heat diffusion
coefficient, $C_p$ is the specific heat per unit mass, 
and $\eta /\rho =1/\beta m\nu $ is the kinematic viscosity.
The equilvalence of $T$, $D_T$, and $\eta/\rho$ is a peculiarity of the
BGK model.
These coefficients are positive so that Eqs.(\ref{mode1}) represent damped
excitations. Corrections to these dispersion relations are of order $k^3$,
describing an expansion that is analytic in $k$ about $k=0$.

For finite shear rate, the modes are more complicated and the behavior at
long wavelengths is qualitatively different. To be more precise, consider
the case of $k\rightarrow 0$ at fixed, finite $a$. It follows directly from (%
\ref{lapdef}) that the hydrodynamic modes have the asymptotic behavior 
\begin{equation}
\omega ^{(i)}({\bf k},a)\rightarrow \left( 
\begin{array}{l}
c_1(a)k^2 \\ 
-\frac 12(1+i\,\sqrt{3})c_2(a)k^{2/3}+\frac 12(1-i\,\sqrt{3}%
)c_3(a)k^{4/3}+c_4(a)k^2 \\ 
-\frac 12(1-i\,\sqrt{3})c_2(a)k^{2/3}+\frac 12(1+i\,\sqrt{3}%
)c_3(a)k^{4/3}+c_4(a)k^2 \\ 
c_2(a)k^{2/3}+c_3(a)k^{4/3}+c_4(a)k^2 \\ 
\left( \eta (a)/\rho \right) k^2
\end{array}
\right) ,  \label{mode2}
\end{equation}
with the coefficients $c_i(a)$ given by 
\begin{equation}
c_1(a)=\left[ \frac{\partial \eta (a)}{\partial T}\frac{\partial
t_{yy}^{(0)}(a)}{\partial n}-\frac{\partial \eta (a)}{\partial n}\frac{%
\partial t_{yy}^{(0)}(a)}{\partial T}\right] \left( m\frac{\partial
t_{yy}^{(0)}(a)}{\partial T}\right) ^{-1},  \label{c1}
\end{equation}
\begin{equation}
c_2(a)=\left[ \frac{2a^2}{3n^2mk_B}\left( \eta (a)+\gamma
_{xy}^{xy}(a)\right) \frac{\partial t_{yy}^{(0)}(a)}{\partial T}\right]
^{1/3},  \label{c2}
\end{equation}
\begin{eqnarray}
c_3(a) &=&\frac 2{9n^2mk_Bc_2(a)}\left[ \frac{\partial t_{yy}^{(0)}(a)}{%
\partial T}(-a\gamma _{yy}^{xy}(a)+t_{yy}^{(0)}(a))-a\,(\eta (a)+\gamma
_{xy}^{xy}(a))\frac{\partial t_{xy}^{(0)}(a)}{\partial T}\right]   \nonumber
\\
&&\hspace{0.2in}+\frac 1{3nmc_2(a)}\left( n\frac{\partial t_{yy}^{(0)}(a)}{%
\partial n}-a\gamma _{xy}^{yy}(a)\right) ,  \label{c3}
\end{eqnarray}
\begin{equation}
c_4(a)=-\frac{c_1(a)}3+\frac 13\left[ \rho ^{-1}\left( \gamma
_{xy}^{xy}(a)+\gamma _{yy}^{yy}(a)\right) +\frac 2{3nk_B}\xi
_{T,y}^y(a)\right] .  \label{c4}
\end{equation}
The modes (\ref{mode2}) represent two oscillating modes and three purely
damped modes, just as in the Navier-Stokes case (\ref{mode1}). However,
there are two important qualitative differences. First, the long wavelength
behavior is non-analytic in $k$ about $k=0$ and is given by a power series
in $k^{2/3}$. Thus, for example the purely damped modes do not represent
diffusive behavior as in the Navier-Stokes case. This non-analytic behavior
with respect to $k$ is due to the fact that the reference matrix at $k=0$ is
not diagonalizable and the eigenvalues are not analytic about $k=a=0$.
Therefore, recovery of the form of the modes near equilibrium requires that $%
k$ and $a$ be taken to zero in a related way (see below). The hydrodynamics
for an alternative choice of thermostat, discussed in Appendix A,  has
dispersion relations that are analytic about $k=a=0$. A second critical
difference between(\ref{mode2}) and (\ref{mode1}) is that the two
oscillating modes are unstable in the long wavelength limit because the
coefficient $c_2(a)$ is positive for all $a$. This means the modes include
excitations that grow in time. Eventually, the deviations $\delta {%
\widetilde{y}}_\alpha $ grow beyond the limitations of the linear equations
and the full non-linear hydrodynamic equations are required to determine
their ultimate values. These will be different from those of the reference
state, representing the fact that the reference state itself is unstable.

\begin{figure}
\caption{The solid line separates the stable (above) and unstable (below)
domains in the $k^* = k \ell$, $a^* = a/\nu$ parameter space. The dotted line 
is the asymptotic form obtained from section \ref{sec:correl},
($x > 2/\protect\sqrt{5}$).}
\label{fig:ka}
\end{figure}
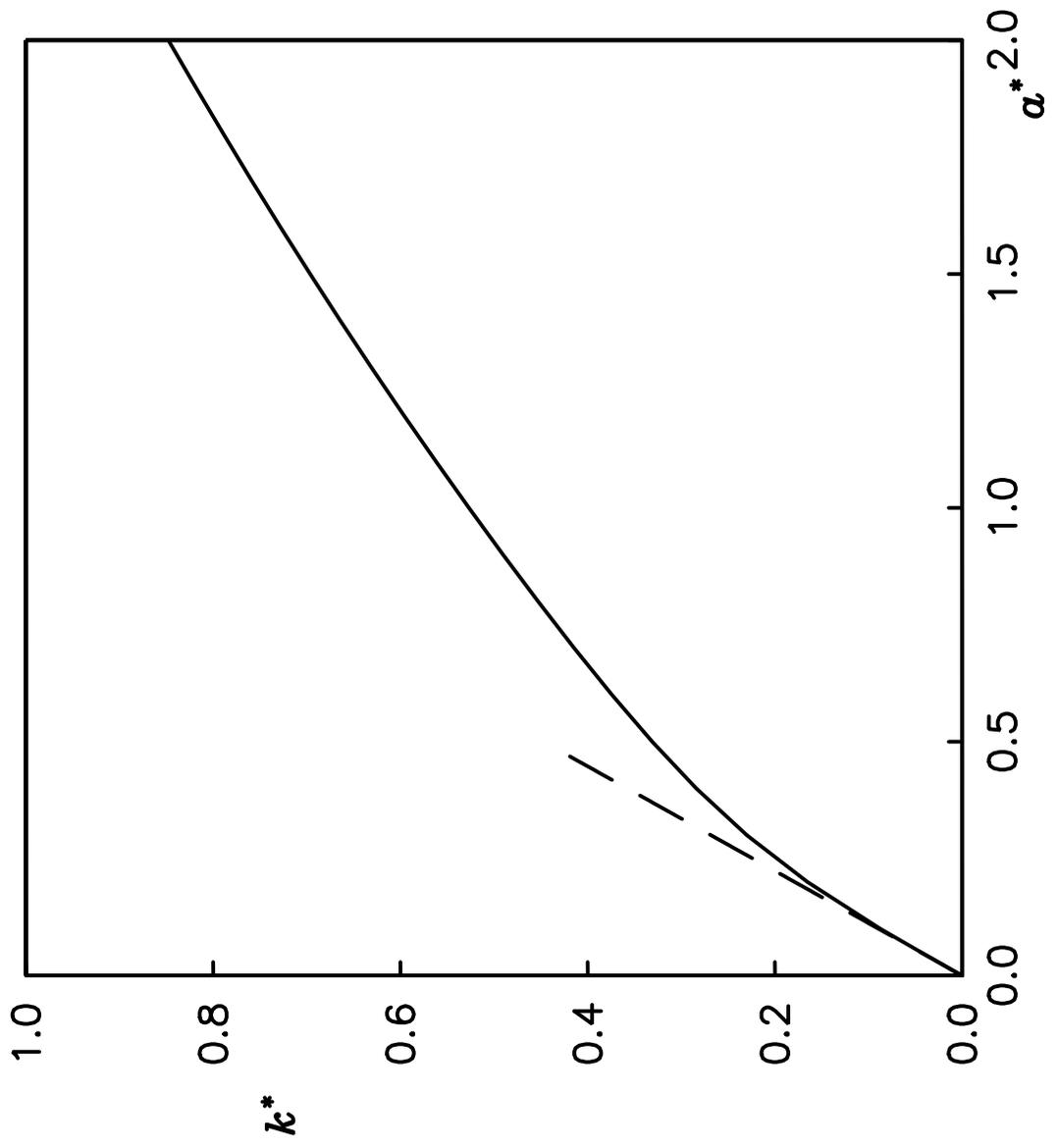

It is possible that the hydrodynamic modes are stable at shorter
wavelengths. This is in fact the case, as can be seen by solving $%
\mathop{\rm Re}[\omega ^{(i)}({\bf k}_{s},a)]=0$ to determine the stability
line $k_s(a)$ in the $k$-$a$ plane. This is illustrated in Figure \ref
{fig:ka} where the calculation was performed using the exact eigenvalues
rather than the small $k$ expansion of (\ref{mode2}). Dimensionless
variables are used, $k^{*}=k\ell $ and $a^{*}=a\tau $, where $\tau =\nu ^{-1}
$ is the mean free time, $\ell =v_0/\nu $ is the mean free path, and $v_0=%
\sqrt{2k_BT/m}$ is the thermal velocity. Above this line the modes are
stable, while below this line they are unstable. This prediction of a long
wavelength instability has been verified quantitatively by comparison with
Monte Carlo simulation of the same kinetic equation from which this
hydrodynamics was derived \cite{mirim:2}. Further analysis of this
instability and comparisons to simulations is reported in detail elsewhere 
\cite{instab}. In the following we focus on the stable domain of 
Fig. \ref{fig:ka}.
To study the stable dynamics a new dimensionless variable, $x=k^{*}/a^{*}$%
, is introduced. The hydrodynamic modes are expressed as functions of $k$
and $x$, i.e., $\omega ^{(i)}({\bf k},a)=$ $\overline{\omega }^{(i)}({\bf k}%
,x)$, and the expansion about $k=0$ is performed at fixed $x$. Physically,
this involves controlling both the shear rate and the wavelength
simultaneously to assure that the system is stable (sufficiently large $x$)
while approaching the long wavelength limit. To simplify the calculation, a
system of Maxwell molecules is considered (interatomic potential $\sim r^{-4}
$). In this case $\nu (n,T)$ and $\lambda (n,T)$ are independent of the
temperature. The hydrodynamic modes for the stable domain are then obtained
for $x>$ $\frac 2{\sqrt{5}}$, which lies above the dashed line shown in
Figure \ref{fig:ka} 
\begin{equation}
\omega ^{(i)}({\bf k},a)\rightarrow \left( 
\begin{array}{c}
ick+\Gamma k^2-\frac{2}{5\nu }a^2 \\ 
-ick+\Gamma k^2-\frac {2}{5\nu }a^2 \\ 
D_Tk^2 \\ 
D_T k^2+\frac {4}{5\nu }a^2  \\ 
\left( \eta /\rho \right) k^2
\end{array}
\right) .  \label{mode3}
\end{equation}
For this expansion at fixed $x$ the eigenvalues are again analytic functions
of $k$ and can be interpreted as perturbations of the Navier-Stokes modes
due to small but finite shear rate. The restrictions on $x$ imply $\frac 2{%
\sqrt{5}}a^{*}<k^{*}<1$, so the shear rate dependence is small but not
necessarily in the Navier-Stokes limit. More generally, the entire stable
domain including larger shear rates can be accessed for $x_s=k/k_s(a)>1$ and
evaluating $\overline{\omega }({\bf k},x_s)$ as a function of $k$ exactly.

The linearized hydrodynamic variables, $\delta \widetilde{y}_\alpha ({\bf k}%
,t)$, can be expressed in terms of the eigenvalues and eigenfunctions, 
\begin{equation}
\delta \widetilde{y}_\alpha ({\bf k},t)=\sum_ie^{-\omega ^{(i)}({\bf k}%
,x)t}\zeta _\alpha ^{(i)}({\bf k},x)(\eta _\beta ^{(i)})^{\dagger}
({\bf k},x)\delta \widetilde{y}_\beta ({\bf k},0),  \label{yeq}
\end{equation}
where $\left\{\zeta^{(i)}\right\} $ are the eigenvectors and 
$\{\eta^{(i)}\}$ are the associated bi-orthogonal set defined by 
$\sum_{\alpha}(\eta^{(i)}_{\alpha})^{\dagger} 
\zeta_\alpha^{(j)}=\delta _{ij}$. 
To illustrate the effects of the
shear rate the analytic results for small $k$, (\ref{mode3}), will be used.
The corresponding eigenvectors are 
\begin{equation}
\zeta ^{(i)}=\left( 
\begin{array}{c}
\left( \sqrt{\frac 35},\,\sqrt{\frac 25},\,\sqrt{\frac 65}\frac ix%
,\,-1,0\right)  \\ 
\left( \sqrt{\frac 35},\,\sqrt{\frac 25},\,\sqrt{\frac 65}\frac ix%
,\,1,0\right)  \\ 
\left( -\sqrt{\frac 23},\,1,\,0,\,0,0\right)  \\ 
\left( -\sqrt{\frac 23},\,1,\,-\frac{2i}{\sqrt{3}x},\,0,0\right)  \\ 
\left( 0,\,0,\,0,\,0,1\right) 
\end{array}
\right) ,
\end{equation}
\begin{equation}
\eta ^{(i)}=\left( 
\begin{array}{c}
\left( \frac 12\sqrt{\frac 35},\,\frac 1{\sqrt{10}},\,0,-\frac 12,0\right) 
\\ 
\left( \frac 12\sqrt{\frac 35},\,\frac 1{\sqrt{10}},\,0,\,\frac 12,0\right) 
\\ 
\left( -\sqrt{\frac 32},\,0,\,\frac{i\sqrt{3}x}2,\,0,0\right)  \\ 
\left( \frac 35\sqrt{\frac 32},\,\frac 35,\,-\frac{i\sqrt{3}x}2,\,0,0\right) 
\\ 
\left( 0,\,0,\,0,\,0,1\right) 
\end{array}
\right) .
\end{equation}
where it is understood that ${\bf k}$ is restricted to the stable domain.
Then the response of the density to an initial density perturbation is found
to be 
\begin{equation}
\delta n({\bf k},t)=\left[\frac{3}{5}e^{-(\Gamma k^2-\frac{2a^2}{5\nu})t}
cos(kct)+e^{-D_T k^2 t}\left(1-\frac35 e^{-\frac{4a^2t}{5\nu}}\right)
\right] \delta n({\bf k},0)
\label{density}
\end{equation}
Thus, for small shear rates there is an enhancement of the 
amplitudes for both the sound and heat modes.
All other hydrodynamic variables can be calculated in a similar
way from (\ref{yeq}) as well for both the stable and unstable regions,
including small and large shear rates.

\section{DISCUSSION}

\label{sec:dis} The objective here has been to study transport far from
equilibrium for the special nonequilibrium states near uniform shear flow.
Dynamical properties of states far from equilibrium are not well understood
due to their complexity and technical difficulties with the formal theories
of nonequilibrium statistical mechanics. However at low density the
Boltzmann kinetic theory provides a controlled formulation of this problem.
There are still difficulties for practical applications so a kinetic model
has been used to allow a detailed analysis for a special class of states
near uniform shear flow. First, the BGK-Boltzmann equation was solved
exactly for the steady state distribution at uniform shear flow, and the
corresponding transport properties given as a function of the shear rate.
Next, a solution to the kinetic equation was obtained for a class of states
deviating from uniform shear flow by small spatial gradients in the
hydrodynamic fields, using a variant of the Chapman-Enskog approximation
method. By ``variant'' is meant that a local form of the stationary solution
for shear flow is used as a reference function rather than the local
equilibrium distribution function. In general this reference state is very
different from a Maxwellian and can be very far from equilibrium if the
shear rate is large. The Chapman-Enskog expansion was used to determine the
distribution function to first order in the gradients. The result is a
''normal solution'' for which all space and time dependence occurs through
the hydrodynamic fields. These fields must be determined from hydrodynamic
equations which follow from the exact conservation laws. The irreversible
parts of the hydrodynamic fluxes were determined as functions of the
hydrodynamic fields and their gradients using the normal solution with
results of the form $t_{ij}^{*}=-\eta \frac{\partial u_{s,i}}{\partial r_j}%
-\gamma _{lk}^{ij}\frac{\partial \delta u_l}{\partial r_k}$ for the momentum
flux and $q_i^{*}=-\xi _{T,j}^i\frac{\partial \delta T}{\partial r_j}-\xi
_{n,j}^i\frac{\partial \delta n}{\partial r_j}$ for the heat flux. Since
these coefficients are calculated near the stationary state of broken
symmetry there are many new transport coefficients ($\gamma _{lk}^{ij}$, $%
\xi _{T,j}^i$, $\xi _{n,j}^i$), in comparison to the case of states near
equilibrium, which depend on the shear rate. With these new expressions for
the fluxes, a closed set of the generalized hydrodynamic equations was
derived near uniform shear flow. In summary, a complete description at both
the kinetic and hydrodynamic levels has been given for a wide class of
states arbitrarily far from equilibrium. In particular, the analysis
provides a rare example of the relevance of a hydrodynamic description far
from equilibrium. The corresponding hydrodynamic modes were calculated to
order $k^2$ for arbitrary shear rate from the linearized hydrodynamic
equations. An unexpected result is an instability for any finite value of
the shear rate at sufficiently long wavelengths. This prediction has been
verified quantitatively by comparison with direct Monte Carlo computer
simulation of the kinetic equation, confirming the validity of the
hydrodynamic description \cite{mirim:2,instab}.

The hydrodynamic analysis was carried out here only for spatial variations
along the velocity gradient. More complex dynamics is expected in the
general case of arbitrary direction for the spatial perturbation and a
description will be given elsewhere. In addition to the stable and unstable
exponential time dependence of the hydrodynamic modes, there will be
algebraic modulations due to the fact that the hydrodynamic matrix at $k=0$
cannot be diagonalized. This can lead to initial growth of perturbations of
uniform shear flow even when the hydrodynamic modes are stable.

The approach taken here can be extended to several other physically
interesting reference non-equilibrium states such as a constant temperature
gradient, or combined heat and momentum transport. Exact solutions to the
BGK kinetic equation are known for these cases \cite{duftyrev} so the
reference distribution for the Chapman-Enskog expansion is available. Other
directions of extension include applications at higher densities. Recently,
a BGK-like kinetic model for the dense fluid Enskog kinetic equation has
been described and applied with success to shear flow \cite{DSB}. This
provides theoretical access to densities relevant for molecular dynamics
simulations.

Although attention has been focused on transport the analysis can be
extended in a straightforward way to describe fluctuations in uniform shear
flow. The reason for this is the close relationship of the kinetic equations
for fluctuations to that for transport \cite{ernst,marchetti:83}. For
example, the kinetic equation for phase space fluctuations at two times is
governed by the linearization of the kinetic equation for transport. As a
consequence, the linear hydrodynamic equations studied in Section 4 also can
be used to compute the hydrodynamic part of time correlation functions such
as the dynamic structure factor measured in light scattering. Recently,
kinetic models for fluctuations have been developed that are self-consistent
with the BGK kinetic model for transport \cite{mirim:1}. Their detailed
application to fluctuations in uniform shear flow, including anomalous long
range spatial correlations \cite{mirim:3} will be discussed elsewhere.

\acknowledgements
The authors wish to thank J.J.Brey, J. Lutsko, and A. Santos for valuable
discussions. This research was supported in part by NSF grant PHY 9312723
and the Division of Sponsored Research at the University of Florida.

\appendix 

\section{Thermostats}

\label{app:const} To establish a steady state for uniform shear flow a
non-conservative external force is introduced in section \ref{sec:kinetic} to
compensate for viscous heating generated by the Lees-Edwards boundary
conditions. In section \ref{sec:hydro} for states {\em near} uniform shear
flow, a local form for this non-conservative force is used 
\begin{equation}
{\bf F}_{ext}({\bf r},t)=-m\lambda (n({\bf r},t),T({\bf r},t))({\bf v}-{\bf u%
}({\bf r},t)).  \label{app:localF}
\end{equation}
With this local form it is possible to impose $\partial _t^{(0)}T=0$, which
led to the results that $\lambda (n({\bf r},t),T({\bf r},t))$ is the same as 
$\lambda _s(n_s,T_s)$ at uniform shear flow given by (\ref{lambda}), except
that the temperature and density are replaced by those for the general
nonequilibrium state. The resulting hydrodynamic equations express viscous
heating only due to the gradients relative to uniform shear flow. An
alternative choice is to use the simpler case of the constant $\lambda
(n_s,T_s)$ in (\ref{app:localF}) even for states near uniform shear flow.
The advantage of this choice is a simpler implementation of the Monte Carlo
simulation method for the solution to the kinetic model equations.
Obviously, it is consistent with the conditions for the steady state uniform
shear flow. However, this choice only partially compensates for the viscous
heating at local uniform shear flow, i.e., $\partial _t^{(0)}T\neq 0$. In
this appendix, the changes from the results of section \ref{sec:hydro} in
both the Chapman-Enskog solution and the hydrodynamic equations due to this
alternative choice are described.

To lowest order in the uniformity parameter the BGK equation with the new
thermostat becomes, 
\begin{equation}
\left( \frac{\partial ^{(0)}}{\partial t}-L(v^{\prime },a)+\lambda
\,_s\delta {\bf u}\cdot {\bf \nabla }_{{\bf v}^{\prime }}\right) f^{(0)}=\nu
f_\ell .  \label{f0eq}
\end{equation}
Here and below it is understood that $\lambda _s=\lambda (n_s,T_s)$ is
independent of space and time. Since the solution is normal, its time
dependence occurs only through the hydrodynamic variables and the
contributions from the time derivative, $\partial _t^{(0)}$, can be
calculated using the conservation laws to lowest order in the uniformity
parameter, 
\begin{equation}
\frac{\partial ^{(0)}n({\bf r},t)}{\partial t}=0,\hspace{0.3in}\frac{%
\partial ^{(0)}\delta u_i({\bf r},t)}{\partial t}+a_{ij}\delta u_j({\bf r}%
,t)=0,
\end{equation}
\begin{equation}
\frac 32n({\bf r},t)k_B\frac{\partial ^{(0)}T({\bf r},t)}{\partial t}%
+a_{ij}t_{ij}^{(0)}({\bf r},t)+3\,\lambda \,_sn({\bf r},t)k_BT({\bf r},t)=0.
\label{newT}
\end{equation}
These differ from the results of the section \ref{sec:hydro} because it is
no longer possible to choose $\lambda _s$ to compensates for the viscous
heating, i.e., $a_{ij}t_{ij}^{(0)}({\bf r},t)\neq -3\,\lambda \,_sn({\bf r}%
,t)k_BT({\bf r},t)$. Instead, Eq.(\ref{f0eq}) becomes, 
\begin{equation}
\left[ -\frac 2{3nk_B}(a_{ij}t_{ij}^{(0)}+3\lambda _s\,nk_BT)\frac \partial {%
\partial T}-L(({\bf v}^{\prime }-\delta {\bf u}),a)\right] f^{(0)}=\nu
\,f_\ell,  \label{f0eq:1}
\end{equation}
where the temperature derivative term is new. To solve the Eq.(\ref{f0eq:1}%
), we need to know the temperature dependence of the zeroth momentum flux, $%
t_{xy}^{(0)}$. For power law potentials this follows from the temperature
dependence of $\nu $. The analysis is simplest for Maxwell molecules where $%
\nu $ is independent of the temperature. Then $t_{xy}^{(0)}\sim T$ (as is
verified a posteriori) and the solution to Eq.(\ref{f0eq:1}) is 
\begin{eqnarray}
&&f^{(0)}({\bf v}^{\prime },y_\alpha ({\bf r},t))=\nu ({\bf r},t)\,n({\bf r}%
,t)\left( \frac m{2\pi k_BT({\bf r},t)}\right) ^{3/2}\int_0^\infty d\tau
\,e^{-\nu ({\bf r},t)\tau }e^{3w({\bf r},t)\tau }  \nonumber \\
&&\quad \quad \quad \exp \left( -\frac m{2k_BT({\bf r},t)}e^{2w({\bf r}%
,t)\tau }[\Lambda _{ij}(-\tau )(v_j^{\prime }-\delta u_j({\bf r}%
,t))]^2\right) ,  \label{newf0}
\end{eqnarray}
\begin{equation}
w(n({\bf r},t),T({\bf r},t))\equiv -\frac{a\,t_{xy}^{(0)}({\bf r},t)}{3n(%
{\bf r},t)k_BT({\bf r},t)}.  \label{w}
\end{equation}
The solution is still only implicit since $t_{xy}^{(0)}({\bf r},t)$ , or
equivalently $w({\bf r},t)$, must be determined self-consistently using (\ref
{newf0}). The result is, 
\begin{equation}
3\,w({\bf r},t)(2w({\bf r},t)+\nu ({\bf r},t))^2=\nu \,({\bf r},t)a^2.
\label{weq}
\end{equation}
This equation is the same as that for $\lambda (n({\bf r},t),T({\bf r},t))$,
Eq. (\ref{lamrel:2}), showing that $w({\bf r},t)$ has the same functional
dependence on $a$, $\nu ({\bf r},t)$, and $T({\bf r},t)$. This proves that
the zeroth order solutions to the BGK-Boltzmann equation, (\ref{newf0}) and (%
\ref{f0}), for the two different thermostats are the same. The lowest order
equations for the temperature differ, however. Use of (\ref{w}) in (\ref
{newT}) gives,

\begin{equation}
\partial _t^{(0)}T({\bf r},t)=2\,(w-\lambda _s)\,T({\bf r},t),
\label{newT:2}
\end{equation}
which vanishes only at the steady state.

Now consider the first order solution in the Chapman-Enskog expansion, 
\begin{equation}
\left( \frac{\partial ^{(0)}}{\partial t}-L(v^{\prime },a)+\lambda
\,_s\delta {\bf u}\cdot {\bf \nabla }_{{\bf v}^{\prime }}\right)
f^{(1)}=-\left( \frac{\partial ^{(1)}}{\partial t}+(v_i^{\prime }+a_{ij}r_j)%
\frac \partial {\partial r_i}\right) f^{(0)},  \label{f1eq}
\end{equation}
and look for solutions of the form, 
\begin{equation}
f^{(1)}({\bf v}^{\prime}, y_\alpha ({\bf r},t))=X_{n,i}({\bf v}^{\prime
}, y_\alpha ({\bf r},t))\frac{\partial n}{\partial r_i}+X_{T,i}({\bf v}%
^{\prime },y_\alpha ({\bf r},t))\frac{\partial T}{\partial r_i}+X_{u_k,i}(%
{\bf v}^{\prime }, y_\alpha ({\bf r},t))\frac{\partial \delta u_k}{\partial
r_i}.  \label{form}
\end{equation}
The equations for the coefficients, $X_{\beta ,i}({\bf v}^{\prime };y_\alpha
({\bf r},t),)$, are determined from substitution of into the (\ref{f1eq})
and they are 
\begin{equation}
\left( \frac{\partial ^{(0)}}{\partial t}-L(v^{\prime },a)+\lambda
\,_s\delta {\bf u}\cdot {\bf \nabla }_{{\bf v}^{\prime }}\right) \left( 
\begin{array}{c}
X_{n,k} \\ 
X_{T,k} \\ 
X_{u_j,k}
\end{array}
\right) +\left( 
\begin{array}{c}
2\,T\left( \frac{\partial w}{\partial n}\right) \,X_{T,k} \\ 
2\,(\lambda _s-w)\,X_{T,k} \\ 
a\,\delta _{j,y}\,X_{u_j,k}
\end{array}
\right) =-\left( 
\begin{array}{c}
Y_{n,k} \\ 
Y_{T,k} \\ 
Y_{u_j,k}
\end{array}
\right) ,  \label{newXeq}
\end{equation}
where $i,j,k$ represent Cartesian coordinate. The functions $Y_{\beta ,k}$
are same as those given in Appendix B. Since the equations for $%
X_{n,k},X_{T,k}$ are different from (\ref{Xeq}), the transport coefficients
for the heat flux, $\xi _{n,j}^i,\xi _{T,j}^i$, are also different for this
new thermostat. However, the equation for $X_{u_i,k}$ is unchanged so the
transport coefficients for the momentum flux, $\gamma _{lm}^{ij}$, are
unchanged. The corresponding hydrodynamic equations are the same as in the
section \ref{sec:hydro}, except for the temperature equation, which becomes 
\begin{eqnarray}
&&\frac 32nk_B\,D_tT-a_{ij}\gamma _{lm}^{ij}\frac{\partial \delta u_l}{%
\partial r_m}+\frac{\partial \delta u_i}{\partial r_j}[t_{ij}^{(0)}-\gamma
_{lm}^{ij}\frac{\partial \delta u_l}{\partial r_m}]  \nonumber \\
&&\quad \quad -\frac \partial {\partial r_i}[\xi _{n,j}^i\frac{\partial n}{%
\partial r_j}+\xi _{T,j}^i\frac{\partial T}{\partial r_j}]=3(w-\lambda
_s)nk_BT.
\end{eqnarray}
The new term, $3\,(w-\lambda _s)nk_BT$, on the right side represents the
viscous heating due to spatially uniform deviations from the steady state.

\section{CHAPMAN ENSKOG EXPANSION}

\label{app:X} In this appendix, Eq.(\ref{first}) for $f^{(1)}$ is solved, 
\begin{equation}
\left( \frac{\partial ^{(0)}}{\partial t}-L(v^{\prime },a)+\lambda \delta 
{\bf u}\cdot {\bf \nabla }_{{\bf v}^{\prime }}\right) f^{(1)}=-\left(
D_t^{(1)}+{\bf v}^{\prime }\cdot {\bf \nabla }\right) f^{(0)},
\label{app:first}
\end{equation}
where $D_t^{(1)}\equiv \frac{\partial ^{(1)}}{\partial t}+a_{ij}r_j\frac 
\partial {\partial r_i}$. The right side of Eq.(\ref{app:first}) can be
written in terms of hydrodynamic derivatives, 
\begin{equation}
(D_t^{(1)}+{\bf v}^{\prime }\cdot {\bf \nabla })f^{(0)}=\sum_{\alpha =1}^5%
\frac{\partial f^{(0)}}{\partial y_\alpha }(D_t^{(1)}+{\bf v}^{\prime }\cdot 
{\bf \nabla })y_\alpha ,  \label{CP:1}
\end{equation}
where $y_\alpha =\{n({\bf r},t),\,T({\bf r},t),\,\delta {\bf u}({\bf r},t)\}$%
. The time derivatives can be replaced by first order spatial gradients of
hydrodynamic variables using the corresponding hydrodynamic equations
obtained from Eqs.(\ref{eqn})-(\ref{equ}) 
\begin{equation}
D_t^{(1)}n+{\bf \nabla }\cdot \left( n\delta {\bf u}\right) =0,
\label{firstn}
\end{equation}
\begin{equation}
\frac 32nk_B(D_t^{(1)}+\delta {\bf u}\cdot {\bf \nabla }%
)T+a_{ij}t_{ij}^{(1)}+t_{ij}^{(0)}\frac{\partial \delta u_i}{\partial r_j}+%
{\bf \nabla }\cdot {\bf q}^{(0)}=0,  \label{firstT}
\end{equation}
\begin{equation}
D_t^{(1)}\delta u_l+\delta {\bf u}\cdot {\bf \nabla }\delta u_l+\rho ^{-1}%
\frac{\partial t_{il}^{(0)}}{\partial r_i}=0.  \label{firstu}
\end{equation}
Use of these equations in Eq.(\ref{CP:1}) gives 
\begin{equation}
(D_t^{(1)}+{\bf v}^{\prime }\cdot {\bf \nabla })f^{(0)}=Y_{n,i}({\bf v}%
^{\prime };y_\alpha ({\bf r},t))\frac{\partial n}{\partial r_i}+Y_{T,i}({\bf %
v}^{\prime };y_\alpha ({\bf r},t))\frac{\partial T}{\partial r_i}+Y_{u_k,i}(%
{\bf v}^{\prime };y_\alpha ({\bf r},t))\frac{\partial \delta u_k}{\partial
r_i}.  \label{CP:2}
\end{equation}
The coefficients, $Y_{\alpha ,i}$, are given by 
\begin{equation}
Y_{n,i}=\frac{\partial f^{(0)}}{\partial n}(v_i^{\prime }-\delta u_i)-\rho
^{-1}\frac{\partial f^{(0)}}{\partial \delta u_l}\frac{\partial t_{il}^{(0)}%
}{\partial n},  \label{Yn}
\end{equation}
\begin{equation}
Y_{T,i}=\frac{\partial f^{(0)}}{\partial T}(v_i^{\prime }-\delta u_i)-\rho
^{-1}\frac{\partial f^{(0)}}{\partial \delta u_l}\frac{\partial t_{il}^{(0)}%
}{\partial T},  \label{YT}
\end{equation}
\begin{equation}
Y_{u_k,i}=-n\frac{\partial f^{(0)}}{\partial n}\delta _{ki}-\frac 2{3nk_B}%
\frac{\partial f^{(0)}}{\partial T}\left( t_{ki}^{(0)}+a_{lm}\frac{\partial
t_{lm}^{(1)}}{\partial \left( \frac{\partial \delta u_k}{\partial r_i}%
\right) }\right) +\frac{\partial f^{(0)}}{\partial \delta u_k}(v_i^{\prime
}-\delta u_i),  \label{Yu}
\end{equation}
where $i,l,k,m$ denote $x,y,z$. The derivatives of $f^{(0)}$ can be obtained
simply from Eq.(\ref{f0}), 
\begin{equation}
\frac{\partial f^{(0)}}{\partial n}=\frac{f^{(0)}}n+\left( \frac{\partial
\nu }{\partial n}\right) \frac 1\nu [f^{(0)}-\nu ^2h_\nu ]+\left( \frac{%
\partial \lambda }{\partial n}\right) \nu h_\lambda ,  \label{fn}
\end{equation}
\begin{equation}
\frac{\partial f^{(0)}}{\partial T}=-\frac 3{2T}f^{(0)}+\frac m{2k_BT^2}\nu
h_T+\left( \frac{\partial \nu }{\partial T}\right) \frac 1\nu [f^{(0)}-\nu
^2h_\nu ]+\left( \frac{\partial \lambda }{\partial T}\right) \nu h_\lambda ,
\label{fT}
\end{equation}
\begin{equation}
\frac{\partial f^{(0)}}{\partial \delta u_l}=\frac m{k_BT}\nu h_{u_l}.
\label{fu}
\end{equation}
Here the functions $h_\alpha $ are defined by, 
\begin{equation}
h_\nu =\int_0^\infty d\tau \,\tau e^{-\nu \tau +3\lambda \tau }f_\ell
(e^{\lambda \tau }\Lambda _{ij}(-\tau )(v_j^{\prime }-\delta u_j)),
\label{hnu}
\end{equation}
\begin{eqnarray}
h_\lambda &=&\int_0^\infty d\tau \,e^{-\nu \tau +3\lambda \tau }\left\{
3\tau -\frac m{k_BT}e^{2\lambda \tau }\tau [\Lambda _{ij}(-\tau
)(v_j^{\prime }-\delta u_j)]^2\right\}  \nonumber \\
&&\quad \quad \quad \times f_\ell (e^{\lambda \tau }\Lambda _{ij}(-\tau
)(v_j^{\prime }-\delta u_j)),  \label{hlam}
\end{eqnarray}
\begin{equation}
h_T=\int_0^\infty d\tau \,e^{-\nu \tau +3\lambda \tau }e^{2\lambda \tau
}[\Lambda _{ij}(-\tau )(v_j^{\prime }-\delta u_j)]^2f_\ell (e^{\lambda \tau
}\Lambda _{ij}(-\tau )(v_j^{\prime }-\delta u_j)),  \label{hT}
\end{equation}
\begin{equation}
h_{u_k}=\int_0^\infty d\tau \,e^{-\nu \tau +3\lambda \tau }e^{2\lambda \tau
}g_k({\bf v}^{\prime },\tau )f_\ell (e^{\lambda \tau }\Lambda _{ij}(-\tau
)(v_j^{\prime }-\delta u_j)),  \label{hu}
\end{equation}
where the functions $g_k({\bf v}^{\prime },t)$ in the expression for $%
h_{u_k} $ are 
\begin{eqnarray}
g_x &=&(v_x^{\prime }-\delta u_x)+at(v_y^{\prime }-\delta u_y),  \nonumber \\
g_y &=&at(v_x^{\prime }-\delta u_x)+(1+a^2t^2)(v_y^{\prime }-\delta u_y), 
\nonumber \\
g_z &=&v_z^{\prime }-\delta u_z.  \label{gs}
\end{eqnarray}
The right side of Eq.(\ref{app:first}) now can be written in terms of the
coefficients, $Y_{\alpha ,i}$, and the hydrodynamic gradients. Note that the 
$Y_{\alpha ,i}$ are given only implicitly since they depend on the unknown
flux $t_{ij}^{(1)}$. The procedure is therefore to determine $f^{(1)}$ in
terms of $t_{ij}^{(1)}$, and then use that result to obtain a
self-consistent equation for $t_{ij}^{(1)}$.

Clearly, $f^{(1)}$ must have the same form as the right side of (\ref{CP:2}) 
\begin{equation}
f^{(1)}=X_{n,i}({\bf v}^{\prime },y_\alpha ({\bf r},t))\frac{\partial n}{%
\partial r_i}+X_{T,i}({\bf v}^{\prime },y_\alpha ({\bf r},t))\frac{\partial T%
}{\partial r_i}+X_{u_k,i}({\bf v}^{\prime },y_\alpha ({\bf r},t))\frac{%
\partial \delta u_k}{\partial r_i}.  \label{form2}
\end{equation}
Substitution of this form into (\ref{app:first}) gives the equations 
\begin{equation}
\left( \frac{\partial ^{(0)}}{\partial t}-L(v^{\prime },a)+\lambda \,\delta
u_i\frac \partial {\partial v_i^{\prime }}\right) X_{\alpha
,k}-aX_{u_x,k}\delta _{\alpha ,u_y}=-Y_{\alpha ,k},  \label{app:Xeq}
\end{equation}
where $\alpha =n,T,u_x,u_y,u_z$. These equations have a form similar to
Eq.(4.8) and can be solved in a similar way to get 
\begin{eqnarray}
X_{\alpha ,k} &=&\int_0^\infty d\tau \,e^{-\nu \tau }e^{3\lambda \tau
}[a\delta _{\alpha ,u_y}X_{u_x,k}(e^{\lambda \tau }\Lambda _{ij}(-\tau
)(v_j^{\prime }-\delta u_j))  \nonumber \\
&&-Y_{\alpha ,k}(e^{\lambda \tau }\Lambda _{ij}(-\tau )(v_j^{\prime }-\delta
u_j))].  \label{Xsol}
\end{eqnarray}
This implicit result is sufficient for determination of the transport
coefficients, as described in the next section.

\section{TRANSPORT COEFFICIENTS}

\label{app:coefs} In this appendix, the transport coefficients defined in
Eqs.(\ref{newtij}) are calculated using the solution to Eq.(\ref{Xeq}).
First, the coefficients from momentum flux are considered, 
\begin{equation}
\gamma _{lm}^{ij}=-\int d{\bf v}^{\prime }mc_ic_jX_{u_l,m},  \label{xyxy:1}
\end{equation}
where ${\bf c}={\bf v}^{\prime }-\delta {\bf u}$. As an example, $\gamma
_{xy}^{xy}$ is analyzed in detail. Substitution of the solution, $X_{u_x,y}$%
, (\ref{Xsol}) into the definition of $\gamma _{xy}^{xy}$, (\ref{xyxy:1})
gives 
\begin{equation}
\gamma _{xy}^{xy}=\int d{\bf v}^{\prime }\,mc_xc_y\int_0^\infty dt\,e^{-\nu
t}e^{3\lambda t}Y_{u_x,y}(e^{\lambda t}\Lambda _{ij}(-t)c_j).
\end{equation}
To proceed it is necessary to know the temperature dependence of $\nu $ and $%
\lambda $. The simplest case is that of Maxwell molecules for which $\nu $
and $\lambda $ are constants. Then $\gamma _{xy}^{xy}$ can be calculated
with a sequence of change of variables: first ${\bf v}^{\prime }\rightarrow 
{\bf v}^{\prime }+\delta {\bf u}$, then ${\bf v}^{\prime }\rightarrow
e^{-\lambda t}{\bf v}^{\prime }$, and finally $v_i^{\prime }\rightarrow
\Lambda _{ij}(t)v_j^{\prime }$. In this way $\gamma _{xy}^{xy}$ is found to
be, 
\begin{eqnarray}
\gamma _{xy}^{xy} &=&\int_0^\infty dt\,e^{-(\nu +2\lambda )t}m\int d{\bf v}%
\,(v_x-atv_y)v_y\left[ -\frac 2{3nk_B}\frac{\partial f^{(0)}}{\partial T}%
(t_{xy}^{(0)}+a_{lm}\gamma _{xy}^{lm})+\frac{\partial f^{(0)}}{\partial
\delta u_x}v_y\right]  \nonumber \\
&=&(t_{xy}^{(0)}+a\gamma _{xy}^{xy})\left( \frac{2\nu a}{(2\lambda +\nu )^3}-%
\frac{10\nu a}{3(2\lambda +\nu )^3}\right) -\frac \nu {(2\lambda +\nu )^2}p.
\label{xyxy:3}
\end{eqnarray}
Use of the result (\ref{txy:1}) for $t_{xy}^{(0)}$ and the relation (\ref
{lamrel:2}) for $\lambda $ gives the final value of the $\gamma _{xy}^{xy}$, 
\begin{equation}
\gamma _{xy}^{xy}=\frac{\nu (\nu -2\lambda )}{(2\lambda +\nu )^2(6\lambda
+\nu )}p.  \label{gxyxy}
\end{equation}
The same method can be applied for all the transport coefficients, $\gamma
_{lm}^{ij}$, $\xi _{n,j}^i$ and $\xi _{T,j}^i$. A complete listing of the
results can be obtained from the authors on request.

\end{document}